\documentclass[aps,twocolumn,prd,showpacs,showkeys,preprintnumbers,superscriptaddress,nobibnotes,notitlepage,floatfix,longbibliography,nofootinbib]{revtex4-2}

\usepackage{graphicx}
\usepackage{amsmath}
\usepackage{amssymb}
\usepackage{xspace}
\usepackage{xcolor}
\usepackage[normalem]{ulem} 

\usepackage[colorlinks=true,citecolor=blue,urlcolor=blue]{hyperref}

\def\be{\begin{equation}}
\def\ee{\end{equation}}
\def\bea{\begin{eqnarray}}
\def\eea{\end{eqnarray}}

\def\mev{\, {\rm MeV}}
\def\kev{\, {\rm keV}}
\def\ev{\, {\rm eV}}

\newcommand{\gsim}{\lower.7ex\hbox{$\;\stackrel{\textstyle>}{\sim}\;$}}
\newcommand{\lsim}{\lower.7ex\hbox{$\;\stackrel{\textstyle<}{\sim}\;$}}

\newcommand{\Neff}{\ensuremath{N_{\mathrm{eff}}}\xspace}
\newcommand{\DNeff}{\ensuremath{\Delta N_\mathrm{eff}}\xspace}
\newcommand{\lcdm}{\ensuremath{\Lambda{\mathrm{CDM}}}\xspace}

\begin{document}

\title{Constraints on the Injection of Radiation in the Early Universe}

\author{Melissa Joseph}
\affiliation{Department of Physics and Astronomy, University of Utah, Salt Lake City, UT  84112, USA}

\author{Jason Kumar}
\affiliation{Department of Physics and Astronomy, University of Hawai'i, Honolulu, HI 96822, USA}

\author{Pearl Sandick}
\affiliation{Department of Physics and Astronomy, University of Utah, Salt Lake City, UT  84112, USA}

\begin{abstract}
We consider the generic injection of radiation (both dark and electromagnetic) during the epoch between 
big bang nucleosynthesis (BBN) and recombination.  The contribution of the additional radiation to the number of 
effective neutrinos may be quite small in this scenario, since dark radiation and electromagnetic radiation 
provide contributions of opposite sign.  However, the injection of electromagnetic radiation dilutes the 
baryon-to-entropy ratio, which is measured both at BBN and at recombination.  As a result, this scenario 
is expected to be tightly constrained.  Indeed, performing a numerical study, we find that the allowed amount of extra  
radiation may be no more than $\sim 25\%$  greater than in the case where it is assumed to be entirely dark radiation.
\end{abstract}

\maketitle

\section{Introduction}

Processes such as early-Universe phase transitions can impact the cosmological evolution of the Universe by 
depositing energy in the form of relativistic particles, which may be either light Standard Model (SM) 
particles or new relativistic degrees of freedom.  These processes are constrained by cosmological observables 
which are sensitive to the radiation energy and entropy densities.  
The effect of these new relativistic degrees of freedom is often characterized by a shift to the number of 
effective neutrinos (\Neff), which characterizes the ratio of the energy density of relativistic particles 
which are not coupled to the photon bath to the energy density of the photons themselves.
As many theories of physics beyond the Standard Model (BSM) predict contributions to \Neff, a key goal of observational cosmology is to constrain 
\DNeff, the deviation of \Neff from the prediction of standard cosmology~\cite{Planck:2018vyg,SimonsObservatory:2018koc,Goldstein:2026iuu}.  

However, this one parameter alone is not sufficient to characterize the impact of 
adding relativistic particles in the early Universe.  Although \DNeff characterizes the impact 
on the energy density of relativistic particles, the impact on the photon entropy density cannot be determined without 
additional information.  
The goal of this work will be to determine, using 
observational cosmology data, how much 
energy can be injected before recombination if we remain agnostic about its composition.

To illustrate the insufficiency of \DNeff as a parameter, one can consider two possibilities.  On the one hand, 
if some beyond-the-Standard-Model (BSM) physics process produces 
relativistic particles which do not couple to the SM, then they will 
provide a positive contribution to  \DNeff.  On the other hand, 
if a BSM process produces 
photons, then the ratio of the neutrino energy density to the photon energy density will decrease, 
yielding a negative contribution to  \DNeff.  Both processes could result from a cosmological 
phase transition, for example, yielding contributions to \DNeff which tend to cancel out.  But the injection of 
photons after big bang nucleosynthesis (BBN) will increase the entropy density of the photon bath relative 
to the baryon number density, an effect which is not captured by the total \DNeff.  A change in the 
late-time baryon-to-entropy density ratio will in turn affect the baryon-to-matter ratio, and could produce tension 
with measurements from the cosmic microwave background (CMB)~\cite{Planck:2018vyg}. 
Indeed, the production of additional entropy after BBN is expected to be significantly constrained (see, for example, 
Ref.~\cite{Escudero:2026mgw}).  Here, we will determine quantitatively the limits on how much entropy and dark 
radiation can be injected before recombination.

For concreteness, we will consider two scenarios.  In the first case, extra energy is present before BBN in the form of a particle which is relativistic and decoupled from the Standard Model thermal bath.  After BBN, this 
particle begins to redshift like matter, and then decays well before recombination to a combination of photons 
and dark radiation.  We will treat this specific scenario as an example of the more general case in which a 
new physics process leads to additional dark radiation before or around BBN, but the nature of this dark radiation is 
unknown.  Rather than assuming it continues to behave as dark radiation all the way to the epoch of recombination, 
we consider the case in which some or all of the dark radiation can essentially convert to photons through decays, 
for example.  The goal is to estimate by how much the cosmological constraints on the injection of dark radiation can 
be shifted or weakened by the details of dark sector particle physics.

The second scenario that we will consider is a cosmological first-order phase transition occurring between BBN and recombination, 
in which both dark sector radiation and photons are released.  As in the first scenario, this case allows us to study the 
constraints on early Universe radiation injection if we remain agnostic to the composition of the energy density.  But in this 
case, the additional energy density does not affect the generation of light element abundances during BBN, since at that time 
there is only an additional vacuum energy density, which is very subleading.

Phase transitions in the dark sector provide a well-motivated physical mechanism for the injection of radiation between BBN and recombination~\cite{Niedermann:2019olb,Niedermann:2020dwg,Garny:2024ums,Greene:2026gnw,Bai:2021ibt,Bringmann:2023opz}.  The cosmological constraints on such scenarios, as well as on radiation injection more generally, have been studied using joint BBN and CMB analyses~\cite{Sobotka:2022vrr,Millea:2015qra,Balazs:2022tjl,Cadamuro:2011fd,Depta:2020wmr,Masso:1995tw,Sabti:2020yrt,Giovanetti:2021izc,Yeh:2022heq}.  Our work complements these studies by remaining agnostic about the composition of the injected energy, and determining how much freedom is gained by allowing a mix of dark and electromagnetic radiation.
Entropy injection between the epochs of BBN and recombination has also been considered 
in Ref.~\cite{Sobotka:2022vrr}, for example.  In that work, however, the authors focused on 
the decays of dark sector particles which redshifted like matter even at the time of BBN.  
Our focus instead is on the injection of particles which redshift like radiation at the 
time of BBN, and which may redshift like matter only over a relatively brief epoch.

We perform a joint Bayesian analysis of BBN and CMB observables\footnote{Note, we do not consider spectral 
distortions of the CMB, because we require that all electromagnetic energy deposition is completed before the epoch in 
which it can distort the CMB spectrum.} using the \texttt{CLASS} Boltzmann solver code and the 
\texttt{LINX} BBN code.
We find that, even remaining agnostic about the composition of the injected radiation, 
there is only a limited amount of extra freedom to be gained.  In fact, the constraints on the extra 
radiation density introduced before BBN are almost as tight as in the case where it is assumed to remain entirely 
dark radiation.  This reflects the tight constraints on the baryon-to-matter ratio both at 
the epoch of BBN (placed by light element abundance measurements) and at the epoch of recombination 
(placed by CMB observations), which limit the amount of entropy injection.  But if extra radiation is introduced 
after BBN as the result of a phase transition, then the allowed extra comoving energy density can be up to 
$25\%$ larger.

The plan of this paper is as follows.  In Section~\ref{sec:GeneralScenario} we  describe the 
scenario of early Universe radiation injection considered here, and its implications for cosmology.  
In Section~\ref{sec:Analysis}, we describe the details of our numerical analysis.
We present our results in Section~\ref{sec:Results}, and conclude in Section~\ref{sec:Conclusion}.

\section{General Scenario}
\label{sec:GeneralScenario}

In the standard cosmology, we can express the energy density ($\rho$) and entropy 
density ($s$) of the Standard Model plasma as 
\bea
\rho  (\tau_{\gamma}) &=& g_{*\rho} (T_{\gamma})~ \frac{\pi^2}{30} T_{\gamma}^4 ,
\nonumber\\
s (\tau_{\gamma}) &=& g_{*s}  (T_{\gamma})~  \frac{2\pi^2}{45} T_{\gamma}^3 ,
\eea
where $T_\gamma$ is the temperature of the photon bath and $g_{*\rho}$ and 
$g_{*s}$ are the numbers of effective relativistic Standard Model energy and entropy degrees of 
freedom, respectively.
If all relativistic particles are thermalized at temperature $T_\gamma$, then 
$g_{*\rho} = g_{*s} = n_B + (7/8) n_F$, where $n_B$ and $n_F$ are the number of 
relativistic bosonic and fermionic degrees of freedom, respectively.

We will begin by considering the radiation-dominated universe, with  $T_\gamma = T_i = 8~\mev$.  At that 
temperature, chosen to be well before BBN, 
the only relativistic SM particles are photons, electrons, positrons and neutrinos.  Below the 
temperature of neutrino decoupling (and before recombination), neutrinos simply redshift as radiation with 
the expansion of the Universe.  But electrons and positrons annihilate away to photons.  This process is 
isentropic, so the comoving entropy density of the electron-photon fluid  ($\propto g_s(T)~ T^3 a^3$)
is conserved.  When electrons and 
positrons annihilate away, the number of effective SM relativistic degrees of freedom (excluding neutrinos) 
changes from $2 + 4(7/8) = 11/2$ to $2$.  As a result, the temperature of the photons must increase by a factor 
of $(11/4)^{1/3}$.  Since the neutrinos are decoupled, their effective temperature $T_\nu$ is unchanged, 
and we find $T_\nu / T_\gamma \approx (4/11)^{1/3}$.  The energy density of the neutrinos may then be 
expressed in terms of the photon temperature as 
\bea
\rho_{\nu} &=& 3 \left(\frac{7}{4} \right) \left(\frac{4}{11} \right)^{4/3} \frac{\pi^2}{30} T_\gamma^4 
= 3 \left(\frac{7}{8} \right) \left(\frac{4}{11} \right)^{4/3} \rho_\gamma ,
\eea
where $\rho_\gamma$ is the energy density of the photons

We may then define the number of effective neutrinos at recombination as 
\bea
N_\mathrm{eff}^\mathrm{CMB} &=& \frac{8}{7} \left(\frac{11}{4} \right)^{4/3} \frac{\rho_\mathrm{rad} - \rho_\gamma}{\rho_\gamma} ,
\eea
where $\rho_\mathrm{rad}$ is the total energy density of all particles which redshift as radiation.  In standard 
cosmology, this definition yields a value $N_{\rm eff}^{\rm SM}$ which is slightly larger than 3, due to some 
physical effects that we have not accounted for (including, for example, that neutrino decoupling 
is not instantaneous~\cite{Mangano:2001iu,Froustey:2020mcq,Akita:2020szl,Bennett:2020zkv}).

If there is some injection of radiation (either SM or dark radiation), then one will obtain a value 
of $\DNeff \equiv \Neff - \Neff^{\rm SM}$ which is non-zero, and may be either positive or negative.  In particular, 
the introduction of dark radiation will increase $\rho_{rad}$ without altering $\rho_\gamma$, yielding a positive 
contribution to \DNeff.  On the other hand, if a new physics process addss photons, then this will 
increase $\rho_\gamma$, yielding a negative contribution to $\Neff$.  This latter case essentially reflects 
the fact that an injection of photons between decoupling and recombination will increase the photon temperature with 
respect to the effective neutrino temperature.

Although the contributions to $\Neff$ from dark radiation and SM radiation tend to cancel, there are 
other effects which cannot be parameterized by \DNeff.  For example, the injection of photons between 
BBN and recombination will shift the entropy density of the photons.  However, the baryon number density 
will not be affected.  This amounts to a shift to the baryon-to-entropy ratio.  The value of this ratio at BBN 
is constrained by measurements of light element abundances, and remains constant in standard cosmology between BBN 
and current epoch.  A shift in the baryon-to-entropy ratio due to the injection of photons after BBN can thus affect 
the baryon density ($\Omega_b$) inferred from BBN and the CMB temperature, potentially creating tension with other CMB observables 
that are sensitive to $\Omega_b$.
So even for a BSM model in which $\DNeff = 0$ at recombination, there may be other effects not parameterized by 
\DNeff which can be constrained by CMB observables.

To study this general scenario, we will consider two specific models.

\subsection{Decay of a light scalar}

We first consider a model in which a relativistic spin-0 
dark sector particle $Y$ is present well before BBN, and thus contributes to $\Neff$ at BBN, and may affect nucleosynthesis.
We will assume that $Y$ subsequently begins to redshift as matter, and then decays to a combination of nearly massless dark radiation and photons.  
We will also assume that these decays occur when the 
photon temperature is $\gtrsim 10~\kev$, 
assuring that the generation of photons from these decays will not distort 
the CMB~\cite{Ota:2014hha,Lucca:2019rxf}.  Note that neither of these assumptions are necessary, but they simplify the analysis by ensuring that some 
constraints from BBN and from CMB distortions are automatically satisfied. 
These assumptions can be weakened, but then 
one would need to check the new constraints in detail.

We can describe this model with three particle physics parameters: the mass of $Y$ ($m$), 
its lifetime ($\tau = 1/\Gamma$), and the branching fraction for decay to photons ($f_\gamma$).  A fourth parameter 
describes the cosmological initial conditions.  We can parameterize the initial conditions with the ratio of 
the dark sector effective temperature to the photon temperature in the epoch before electron-positron decoupling 
($\alpha = T_Y / T_\gamma$).  

The background cosmological evolution is governed by the coupled system of Boltzmann equations for the energy densities and number densities of the various species:
\bea
a \frac{d \rho_Y}{d a} + 3(\rho_Y + P_Y) &=& -m n_Y \frac{\Gamma}{H}, \nonumber\\
a \frac{dn_Y}{d a} + 3 n_Y &=& - m \frac{\Gamma}{H} \int \frac{d^3 p}{(2 \pi)^3 E} f(p), \nonumber\\
a \frac{d \rho_\gamma}{d a} + 4\rho_\gamma &=& m n_Y f_\gamma \frac{\Gamma}{H}, \nonumber\\
a \frac{d \rho_\mathrm{dr}}{d a} + 4\rho_\mathrm{dr} &=& m n_Y (1 - f_\gamma) \frac{\Gamma}{H},
\label{eq:Boltzmann}
\eea
where $a$ is the scale factor, $P_Y$ is the pressure of species $Y$, $n_Y$ is its number density, $\Gamma$ is the decay rate, $H$ is the Hubble parameter, and $f(p)$ represents the phase-space distribution function.\footnote{$n_Y = \int d^3p / (2\pi)^3~ f(p)$.} 
These equations capture the
standard redshift behavior (left-hand sides) augmented by source terms from $Y$  decays (right-hand sides). The decay products partition energy according to the branching fraction $f_\gamma$ with photons and dark radiation receiving fractions $f_\gamma$ and $(1 - f_\gamma)$ respectively.

Light element abundances are affected by the change in the total radiation density during the BBN epoch, which can be expressed in terms 
of the effective number of neutrinos at BBN  as 
\bea
\Delta N_\mathrm{eff}^\mathrm{BBN} &=&  \frac{4}{7} \alpha^4 . 
\label{dneff_bbn}
\eea

CMB observables are related to the change in the comoving energy density of photons and dark radiation due to 
cosmological evolution between BBN and recombination. 
To quantify this, we can define an initial time $t_i$ before BBN, when $Y$ is relativistic.  We will conventionally 
take $t_i$ to be when the SM temperature is $8\mev$.  We similarly define a time $t_f$ after $Y$ has decayed, but still 
well in the radiation-dominated epoch.  We can take $t_f$ to be the time when the photons are at temperature $10~\kev$.
We can then define the ratio of the comoving radiation energy density at late and early times by: 
\bea
r \equiv \frac{a_f^4 \rho_r^f}{a_i^4 \rho_r^i} .
\eea
$r$ quantifies how much the comoving radiation energy density has increased due to electron-positron annihilation 
and  the fact that $Y$ may redshift like matter between the time it becomes non-relativistic and the time it decays. 

When $Y$ decays, the energy density of the photon bath and the dark radiation will change.
The ratios of comoving energy densities at $t_i$ and $t_f$
for photons and ultra-relativistic particles (that is, particles which remain relativistic even up to the 
present epoch), respectively, are:

\bea
r_\gamma &\equiv& \frac{a_f^4 \rho_\gamma^f}{a_i^4 \rho_\gamma^i},
\nonumber\\
r_\mathrm{ur} &\equiv& \frac{a_f^4 \rho_\mathrm{ur}^f}{a_i^4 \rho_\mathrm{ur}^i} = 
\frac{a_f^4}{a_i^4} \frac{ 2\rho^f_{1\nu} + \rho^f_\mathrm{dr}}{ 2\rho^f_{1\nu} + \rho^i_Y} ,
\eea
where $\rho_{1\nu}$ is the energy density of one neutrino species; in this epoch the massless and massive neutrino species have the same energy density.
Ultimately, we are interested in how much extra energy density can be injected, beyond 
what is already provided by Standard Model degrees of freedom in a $\Lambda$CDM cosmology.  We can 
quantify this by defining the quantity $r_\mathrm{SM}$ as 
\bea
r_\mathrm{SM} &=&  \frac{a_f^4 \rho_r^f}{a_i^4 \rho_\mathrm{SM}^i} ,
\eea
where $\rho_\mathrm{SM}^i$ is the density of relativistic SM degrees of freedom at time $t_i$.
Our goal will be to determine how large $r_\mathrm{SM}$ can be, consistent with cosmological observations.

Although we will compute the quantities $r$, $r_\gamma$, $r_\mathrm{ur}$ and $r_\mathrm{SM}$ numerically by 
solving the Boltzmann equations, it will be instructive to derive simple analytic expressions for them in terms 
of the model parameters, having made some simplifying approximations.  We present these approximations in 
Appendix~\ref{app:r_derivation}.

The shift in the ratio of dark radiation density to photon radiation density  
can be parameterized as a change to the number of effective 
neutrinos at recombination, relative to that at BBN.  We can distinguish between neutrinos which 
are effectively massless at current epoch (that is, their mass is small compared to $T_{\gamma 0} \sim 1.3 \times 10^{-4}~\ev$), and 
neutrinos which became non-relativistic between recombination and now.  We assume that, in the Standard Model, there is one massive neutrino 
($m_\nu = 60~\mathrm{meV}$) and 
two ultra-relativistic neutrinos.  

The dark radiation behaves similarly to ultra-relativistic neutrinos, so 
the effective number of massless neutrinos is given by 
\bea
N_\mathrm{ur}^\mathrm{CMB} = \frac{8}{7} \left(\frac{11}{4} \right)^{4/3} \frac{2\rho_{1\nu} + \rho_\mathrm{dr}}{\rho_\gamma}.
\eea

One can parameterize the energy density of the massive neutrino in terms of 
the number of effective massive neutrinos with mass $m = 60~\mathrm{meV}$ ($N_\mathrm{ncdm}$).
This quantity parameterizes the energy density of matter which 
free-streams, and suppresses the growth of structure (which in turn suppresses the 
weak lensing of the CMB). Between BBN and recombination, $N_\mathrm{ncdm}$ is affected only by the change in the comoving energy density of the photons:
\bea
N_\mathrm{ncdm}^\mathrm{CMB} = \frac{8}{7} \left(\frac{11}{4} \right)^{4/3} \frac{\rho_{1\nu}}{\rho_\gamma}  .
\eea
Any injection of energy into the photon bath will decrease $N_\mathrm{ncdm}$ relative to its standard value for a fixed neutrino mass.

If a light scalar couples to photons, then there are potentially a variety of laboratory and astrophysical  
constraints, including constraints from excess supernovae cooling (see, for example,~\cite{Chang:2018rso}).  
However, these constraints are specific to the detailed particle physics model, and are 
thus beyond the scope of this work.

\subsection{First-order dark sector phase transition between BBN and recombination}

We will also consider an alternative model for early Universe cosmology in which there is a
first-order phase transition between the epochs of BBN and recombination.  In this model, the 
phase transition releases latent heat in the form of both dark and electromagnetic radiation.  
The cosmology of this model is similar to that of the decaying scalar model after the scalar has 
decayed completely away, leaving an additional contribution to the dark radiation and electromagnetic 
radiation densities.  Our goal is again to estimate how large $r_\mathrm{SM}$ can be, consistent with 
other observables.

Note that $\rho_\mathrm{SM}^i$ is the same in this scenario as in the scenario in which a scalar $Y$ is 
injected.
But a key difference between these scenarios lies in the generation of light element abundances at BBN.
In the case of a decaying scalar $Y$, the scalar was assumed to be present during BBN and redshifting as 
radiation.  This provided an extra dark radiation density during BBN which could effect the generation of 
the light element abundances.  In the case of a first order phase transition after BBN, the additional 
energy density redshifts as vacuum energy before the transition.  Since radiation dilutes as $a^{-4}$ due 
to cosmological expansion, while vacuum energy density remains constant, we see that the electromagnetic radiation 
density at BBN would be much larger than at the time of the phase transition, and the vacuum energy density at 
BBN would be negligible by comparison.  We thus expect that, if anything, constraints on this scenario 
may be weaker than for the model in which a scalar $Y$ decays.

\section{Analysis}
\label{sec:Analysis}

We perform a comprehensive joint Bayesian analysis of Big Bang Nucleosynthesis (BBN) and 
Cosmic Microwave Background (CMB) observables to constrain the parameters of our decaying 
particle model. Our analysis integrates the Boltzmann solver \texttt{CLASS} 
~\cite{Lesgourgues:2011rh} for CMB power spectrum calculations with a modified version of the BBN code 
\texttt{LINX} (Light Isotope Nucleosynthesis with JAX) ~\cite{Giovanetti:2024zce}, which uses methods and tables from Refs.~\cite{Escudero:2018mvt,EscuderoAbenza:2020cmq,Pitrou:2018cgg,Burns:2023sgx}.

To ensure consistency between the BBN and CMB calculations, the helium mass 
fraction $Y_P$ predicted by \texttt{LINX} at each point in parameter space is passed 
directly as input to \texttt{CLASS} for the CMB power spectrum calculation. This captures 
the impact of $Y_P$ on the CMB damping tail. This approach ensures that both the BBN and CMB likelihoods are evaluated at precisely the same cosmological parameters, with full consistency between the BBN prediction and the CMB calculation.

We additionally implement corrections for the modified neutrino temperature resulting from 
entropy injection. When the $Y$ particle decays to photons, the photon temperature increases relative to the neutrino temperature, altering the standard ratio $T_\nu/T_\gamma = (4/11)^{1/3}$ established by electron-positron annihilation as detailed in Section~\ref{sec:GeneralScenario}. We implement the corrected neutrino temperature, computed by \texttt{LINX}, and correspondingly the updated contribution to $N_\mathrm{ur}$ and $N_\mathrm{ncdm}$, in \texttt{CLASS}. 

Our analysis explores the parameter space encompassing both standard 
cosmological parameters and our model-specific decay parameters. Parameter estimation is performed using the nested sampling algorithm~\cite{Skilling:2004,Skilling:2006,Higson:2018cwj} as implemented in the \texttt{dynesty} package~\cite{Speagle:2020,Koposov:2023}. We use the dynamic nested sampling mode with 500 live points for joint CMB+BBN analyses, and sampling terminates when the estimated contribution of the remaining prior volume to the total evidence 
satisfies $\Delta \ln \mathcal{Z} < 0.5$. The base $\Lambda$CDM parameter set consists of six parameters: the physical baryon density $\omega_b \equiv \Omega_b h^2$, the physical cold dark matter density $\omega_c \equiv \Omega_c h^2$, the angular size of the sound horizon at last scattering $100\theta_s$, the amplitude of the primordial scalar power spectrum $A_s$ (or equivalently $\ln(10^{10} A_s)$), the scalar spectral index $n_s$, and the optical depth to reionization $\tau_{\rm reio}$. For the neutrino sector, we adopt a minimal mass hierarchy consisting of two massless neutrino species and one massive species with $m_\nu = 0.06~\ev$. 

Additionally, to solve the Boltzmann equation, we need the mass $m$ and lifetime $\tau = \Gamma^{-1}$ of $Y$.
But we sample over the more cosmologically relevant parameters $T_\gamma^\mathrm{NR}$ 
and $T_\gamma^\mathrm{decay}$, which roughly parameterize the photon bath temperatures at which $Y$ becomes non-relativistic 
and at which $Y$ decays. 
We define $T_\gamma^\mathrm{NR} = m / \alpha$.  We also define $T_\gamma^\mathrm{decay}$ to be the photon temperature at which the 
age of the Universe is $\tau = \Gamma^{-1}$.
We then solve the Boltzmann equation (eq.~\ref{eq:Boltzmann}) using \texttt{LINX}.

The model-specific parameters characterizing the decaying particle $Y$ and its cosmological 
evolution are:
\begin{itemize}
    \item $\Delta N_{\rm eff}^{\rm BBN}$: the contribution to the effective number of 
    neutrinos at BBN, which parameterizes the initial energy density of the dark sector 
    relative to the Standard Model radiation,
    \item $T_\gamma^{NR}$ [MeV]: the photon temperature when $Y$ transitions from 
    radiation-like to matter-like behavior, 
    \item $T_\gamma^{NR}/T_\gamma^{\rm decay}$: the ratio characterizing the duration of the 
    matter-like epoch and the time-scale over which $Y$ decays, 
    \item $f_\gamma$: the branching fraction for decay to photons (versus dark radiation).
\end{itemize}
We refer to this scenario as the ``decay model."

\begin{figure}
    \centering
    \includegraphics[width=1.\linewidth]{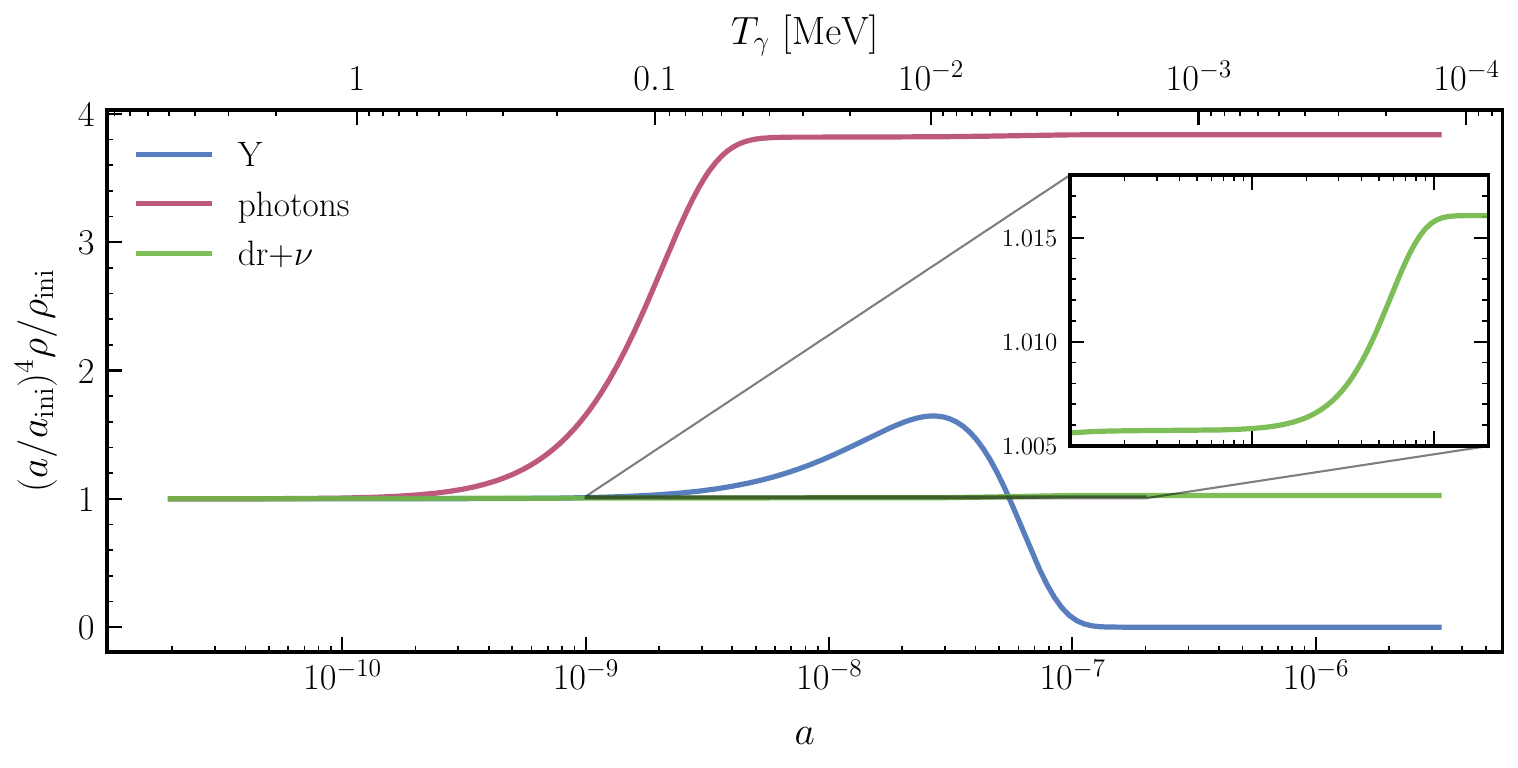}
    \caption{ Background evolution of energy density components for an illustrative decay scenario with parameters $\Delta N_{\rm eff}^{\rm BBN} = 0.015$, $T_{\rm NR} = 0.05 $ MeV, $T_{\rm NR}/T_\gamma^{\rm decay} = 8$, and $f_\gamma = 0.5$. Energy densities of the decaying species $Y$ (blue curve), photons (pink curve), and dark radiation + massless neutrinos (green curve) normalized to their initial energy density. The $Y$ particle begins scaling like matter ($\rho_Y \propto a^{-3}$) at $a \sim 10^{-9}$ until decay starts near $a \sim 10^{-7}$. }
    \label{fig:background_evolution}
\end{figure}

This parameterization is chosen to directly connect to observable quantities: 
$\Delta N_{\rm eff}^{\rm BBN}$ is constrained by light element abundances, $T_\gamma^\mathrm{NR}$ 
determines when the energy density in $Y$ begins to grow relative to radiation, 
$T_\gamma^\mathrm{NR}/T_\gamma^{\rm decay}$ controls the magnitude of this growth (parameterized by $r$ in 
Sec.~\ref{sec:GeneralScenario}), and $f_\gamma$ determines the partitioning between SM and dark sector energy injection.
We plot a solution to the Boltzmann equations for a particular benchmark model in Figure~\ref{fig:background_evolution}.

We adopt flat (uniform) priors on all parameters within physically motivated ranges:
\bea
\omega_b &\in& [0.005, 0.1], \nonumber\\
\omega_c &\in& [0.001, 0.99], \nonumber\\
100\theta_s &\in& [0.5, 10.0], \nonumber\\
\ln(10^{10} A_s) &\in& [1.61, 3.91], \nonumber\\
n_s &\in& [0.8, 1.2], \nonumber\\
\tau_{\rm reio} &\in& [0.01, 0.8], \nonumber\\
\Delta N_{\rm eff}^{\rm BBN} &\in& [0, 2.0], \nonumber\\
T_\gamma^{\rm NR}~[{\rm MeV}] &\in& [0.1, 1], \nonumber\\
T_\gamma^{\rm NR}/T_\gamma^\mathrm{decay} &\in& [1.0, 10.0], \nonumber\\
f_\gamma &\in& [0.0, 1.0].
\eea

The constraint $T_\gamma^\mathrm{NR}/T_\gamma^\mathrm{decay} \geq 1$ enforces the physical requirement that $Y$ becomes non-relativistic before or at the time of decay. The lower bound on $T_\gamma^{\rm decay}$ (corresponding to the upper bound on $T_\gamma^\mathrm{NR}/T_\gamma^\mathrm{decay}$ for the minimal value of $T_\gamma^\mathrm{NR}$) is set to $\gsim 10~\kev$ to ensure that all relevant CMB modes with $\ell \lsim 2500$ enter the horizon after the decay is complete, allowing the use of standard adiabatic initial conditions in \texttt{CLASS}. 

We also perform an analysis for the phase transition scenario described in Section~\ref{sec:GeneralScenario}. 
We will refer to this as the ``PT model."
The main distinction from the decay scenario is that the phase transition does not contribute additional dark radiation during BBN. Since the radiation energy density at BBN is much larger than at the time of a post-BBN phase transition, the vacuum energy contribution at BBN is negligible compared to the radiation density. Therefore, we set $\Delta N_{\rm eff}^{\rm BBN} = 0$ for the phase transition scenario, and the light element abundances are determined solely by the standard BBN expansion history. We use the same priors on the $\Lambda$CDM parameters and additionally require that the phase transition occurs after BBN is complete and that any released photons can thermalize before recombination without producing observable CMB spectral distortions, giving a range $[10, 100]~\kev$ for the phase transition to occur. The additional parameters in this model and their priors are 
$r_\gamma \in [\left( 11/4 \right)^{4/3}, 19]$ and $r_\mathrm{ur} \in [1, 5]$.  

Finally, as a point of comparison for how much additional energy density can be injected at early times as a result of particle physics freedom 
of the dark sector, we analyze the scenario in which, in addition to $\Lambda$CDM, we introduce, before BBN,  dark radiation which cannot decay.  
This latter scenario can be 
considered as a constrained version of our model in which $T_\gamma^\mathrm{NR}/ T_\gamma^{\rm decay} = 1$ 
and $f_\gamma =0$ (so $Y$ redshifts like dark radiation except for a short amount of time, 
and then decays entirely to dark radiation).  We will refer to this model as ``$\Delta N_\mathrm{eff}$".  
We will then see if allowing a more complicated dark sector (i.e., the decay or PT models) allows for the injection of a larger energy 
density than in the standard $\Delta N_\mathrm{eff}$ model.

\subsection{Data}
\subsubsection{CMB Anisotropy Constraints}

Our analysis incorporates temperature and polarization anisotropy measurements from the \textit{Planck} 2018 data release~\cite{Planck:2019nip}. We utilize the full likelihood pipeline consisting of three distinct components: high multipole coverage through the Plik TTTEEE likelihood spanning $30 \leq \ell \leq 2508$, complemented by the Commander temperature likelihood and SimAll polarization likelihood for $2 \leq \ell < 30$. We apply standard \textit{Planck} collaboration priors to these nuisance parameters and perform full marginalization throughout our parameter estimation.

\subsubsection{BBN Observable Constraints}

Our BBN likelihood construction relies on observed primordial abundances of light elements produced during the nucleosynthesis epoch. We incorporate two principal observables that provide complementary constraints on early universe cosmology: the primordial $^4$He mass fraction and the deuterium-to-hydrogen number ratio.  Theoretical predictions for these abundances are computed using \texttt{LINX} with the PRIMAT nuclear reaction network~\cite{Pitrou:2018cgg}.

The $^4$He mass fraction measurement~\cite{Aver:2015iza} provides\footnote{This result is consistent with a very 
recent precise measurement from Ref.~\cite{Aver:2026dxv}.}:
\be
Y_P^{\rm obs} = 0.2449 \pm 0.0040.
\ee

The deuterium abundance from absorption line systems in high-redshift quasars~\cite{Cooke:2017cwo} yields:
\be
({\rm D/H})^{\rm obs} = (2.527 \pm 0.030) \times 10^{-5}.
\ee

\section{Results}
\label{sec:Results}

\begin{figure}
    \centering
    \includegraphics[width=0.99\linewidth]{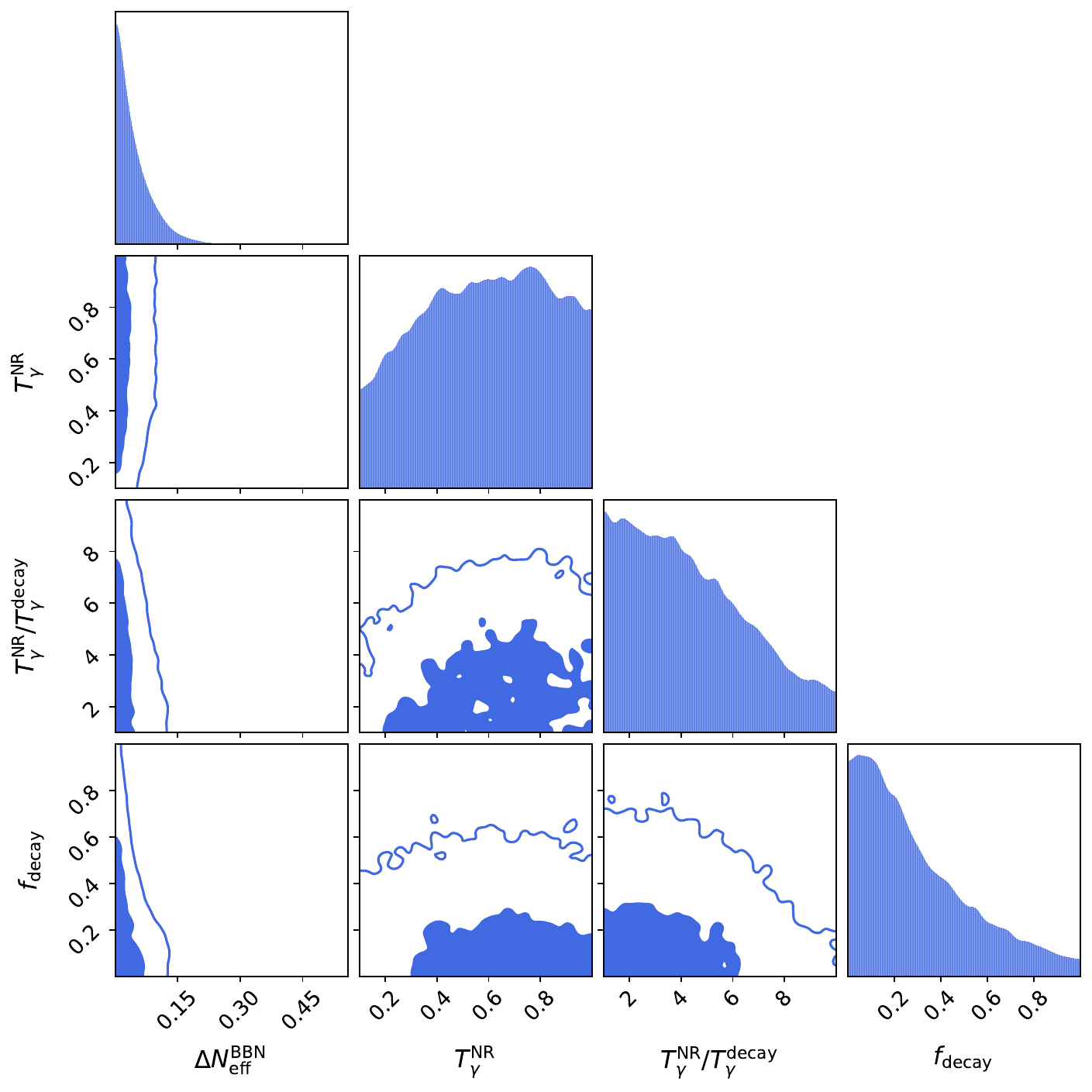}
    \caption{Posterior distributions for the decay model parameters 
    $\Delta N_{\rm eff}^{\rm BBN}$, $T_\gamma^{\rm NR}$, $T_\gamma^{\rm NR}/T_\gamma^{\rm decay}$ and $f_\gamma$, obtained from a joint analysis of \textit{Planck} 2018 CMB temperature and polarization anisotropies combined with primordial $^4$He and D/H abundance measurements. BBN predictions are computed using \texttt{LINX} with the PRIMAT nuclear reaction network. }
    \label{fig:cmb_bbn_decay_results}
\end{figure}

\begin{figure}
    \centering
    \includegraphics[width=0.99\linewidth]{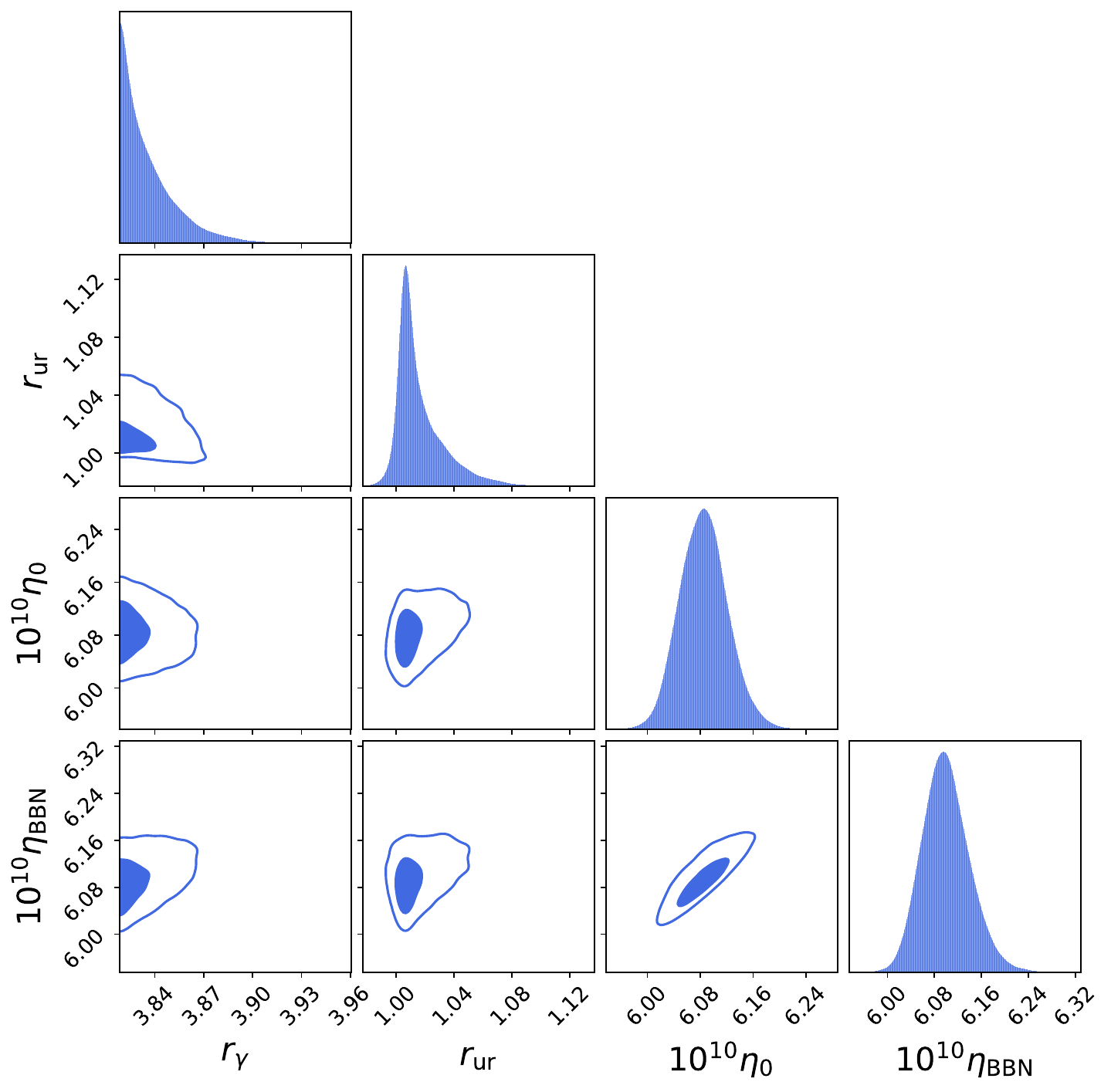}
    \caption{Posterior distributions for derived radiation energy density parameters and the baryon-to-photon ratio at BBN ($\eta_\mathrm{BBN}$) and at the CMB ($\eta_0$), for the decay model. }
    \label{fig:cmb_bbn_decay_r_eta}
\end{figure}

\begin{figure}
    \centering
    \includegraphics[width=0.99\linewidth]{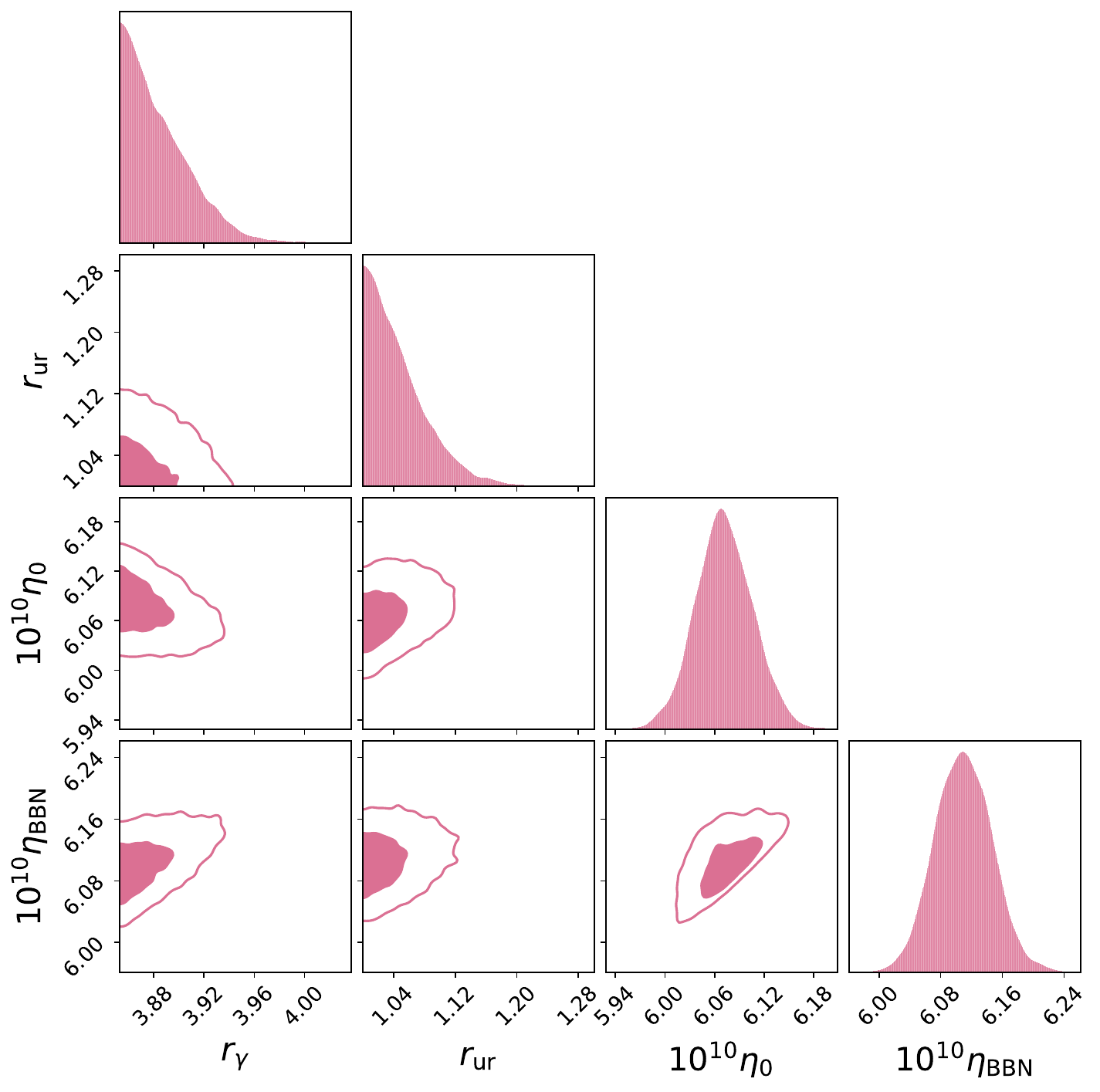}
    \caption{Posterior distributions for the radiation energy density parameters 
    and the baryon-to-photon ratio at BBN ($\eta_\mathrm{BBN}$) and at the CMB ($\eta_0$), for the PT model. Compared to the decay model (Fig.~\ref{fig:cmb_bbn_decay_r_eta}), the PT model allows a marginally larger total injected radiation density, reflected in the slightly looser constraint on $r_\gamma$ and $r_\mathrm{ur}$.  }
    \label{fig:cmb_bbn_PT_r_eta}
\end{figure}

For the decay model, we present the posterior distributions of the model parameters in 
Figure~\ref{fig:cmb_bbn_decay_results}, and of the derived energy density 
ratios and baryon-to-entropy ratios in Figure~\ref{fig:cmb_bbn_decay_r_eta}.  
The posterior distributions of the energy density 
ratios and baryon-to-entropy ratios for the PT model are presented in Figure~\ref{fig:cmb_bbn_PT_r_eta}.
These results are summarized in Table~\ref{tab:Posteriors}.  A more comprehensive set of posterior distributions 
for the $\Delta N_\mathrm{eff}$ model, the decay model and the PT model are presented in Appendix~\ref{sec:CompletePosteriors}.

To study how much additional radiation energy density can be injected, consistent with observation, 
if we remain agnostic about the details of the dark sector, we can consider constraints on the 
parameter $r_\mathrm{SM}$.  This is the ratio of the total comoving radiation energy density when the 
photons are at temperature $10 \kev$ to the comoving energy density of SM degrees of freedom when the photons 
are at temperature $8 \mev$.  This ratio thus increases if additional radiation is injected.  To put these 
results in context, we note that in the standard $\Lambda$CDM cosmology with no additional radiation, we 
would have $r_\mathrm{SM} \sim 1.205$, due to the increase in the comoving radiation density between the 
temperatures of $8 \mev$ and $10 \kev$ as a result of electron-positron annihilation.  A value of $r_\mathrm{SM}$ exceeding 
this would indicate the presence of additional radiation from some other source.

In comparing the posteriors of the $\Delta N_\mathrm{eff}$ and decay models in Table~\ref{tab:Posteriors}, we see that  
allowing a more complicated dark sector through decays of $Y$ allows almost no additional 
freedom for extra radiation at recombination (as measured by $r_\mathrm{SM}$).  The reason is because the addition of electromagnetic 
radiation would dilute the baryon-to-entropy ratio between the epochs of BBN and recombination.  However, 
the baryon-to-entropy ratio can be tightly constrained both from light element abundances (constraining the 
ratio at BBN) and from the CMB (constraining the ratio at recombination).  Indeed, although the data is consistent with a constant baryon-to-entropy ratio, it, if anything, favors a slightly lower value at BBN than at recombination.  This mild tension is related to a well-known discrepancy in the primordial deuterium abundance: depending on the nuclear reaction network employed, the theoretically predicted D/H at the CMB-preferred value of $\omega_b$ can be in up to $\sim 1.8 \sigma$ tension with the measured value~\cite{Pitrou:2020etk,Yeh:2020mgl,Pisanti:2020efz,Giovanetti:2024zce}.  In particular, the PRIMAT nuclear network predicts a D/H abundance that, 
at the baryon density preferred by the CMB, is somewhat higher than the 
observed value, which is equivalent to a preference for a slightly lower baryon-to-entropy ratio at BBN relative to its value inferred from the CMB.  This further disfavors the injection of entropy into the electromagnetic sector after BBN, which would only worsen this discrepancy~\cite{Pitrou:2020etk,Yeh:2020mgl,Pisanti:2020efz}.  

However, comparing the posteriors of the PT model, we note that a slightly larger extra radiation energy density 
(by about $\sim 25\%$) can be introduced in this case than in either the $\Delta N_\mathrm{eff}$ or decay models.  This 
small amount of additional freedom can be traced to the fact that in the PT model, unlike the 
$\Delta N_\mathrm{eff}$ or decay models, the extra dark radiation density present at BBN is negligible.  As such, there is 
no additional distortion of light element abundances arising from a larger Hubble parameter.
To place this in context, we note that in the $\Delta N_\mathrm{eff}$ model, the extra radiation energy density (which 
simply redshifts as dark radiation from temperatures of $8~\mev$ to $10~\kev$) is bounded to be $\Delta N_\mathrm{eff}^\mathrm{BBN} 
< 0.234$ at $95\%$CL.  The additional comoving radiation energy which may be present at $10~\kev$ in the PT model is $\lesssim 25\%$ 
of that.

\begin{table}[h!]
\centering
\begin{tabular}{lccc}
\hline
\textbf{Parameter} & \textbf{$\Delta N_{\rm eff}$} & \textbf{Decay model} & \textbf{PT model} \\
\hline
$10^9 A_s$ 
& $2.0971^{+0.0347}_{-0.0327}$ 
& $2.0983^{+0.0289}_{-0.0293}$ 
& $2.0958^{+0.03353}_{-0.03236}$ \\

$n_s$ 
& $0.9629^{+0.00631}_{-0.00694}$ 
& $0.9630^{+0.00482}_{-0.00431}$ 
& $0.9621^{+0.00480}_{-0.004599}$ \\

$h$ 
& $0.6711^{+0.01064}_{-0.01145}$ 
& $0.6718^{+0.00699}_{-0.00590}$ 
& $0.6706^{+0.00742}_{-0.00665}$ \\

$100 \omega_b$ 
& $2.2230^{+0.01842}_{-0.01836}$ 
& $2.2223^{+0.01288}_{-0.01281}$ 
& $2.2176^{+0.01291}_{-0.01205}$ \\

$\omega_c$ 
& $0.1214^{+0.00199}_{-0.00203}$ 
& $0.12150^{+0.00165}_{-0.00138}$ 
& $0.12211^{+0.00189}_{-0.00155}$ \\

$100 \tau_\mathrm{reio}$ 
& $5.2229^{+0.72440}_{-0.66695}$ 
& $5.2511^{+0.6546}_{-0.6405}$ 
& $5.1249^{+0.74766}_{-0.68615}$ \\

$\Delta N_\mathrm{eff}^\mathrm{BBN}$ 
& $0.0305^{+0.120}_{-0.122}, $ 
& $< 0.133 \left(95\%\right)$ 
& --- \\

& $<0.234 \left(95\%\right)$

\\
$N_\mathrm{ur}^\mathrm{CMB}$ 
& $2.0632^{+0.1259}_{-0.1279}$ 
& $2.0850^{+0.1949}_{-0.0539}$ 
& $2.0921^{+0.0942}_{-0.0894}$ \\

$N_\mathrm{eff}^\mathrm{CMB}$ 
& $3.0764^{+0.1260}_{-0.1279}$ 
& $3.0897^{+0.1801}_{-0.0488}$ 
& $3.0975^{+0.0947}_{-0.1021}$ \\

$T_\gamma^\mathrm{NR}\,[\mathrm{MeV}]$ 
& --- 
& $0.594^{+0.264}_{-0.289}$ 
& --- \\

$T_\gamma^\mathrm{NR}/T_\gamma^\mathrm{decay}$ 
& --- 
& $< 8.732 \left(95\%\right)$ 
& --- \\

$f_\gamma$ 
& --- 
& $< 0.73 \left(95\%\right)$ 
& --- \\

$r_\gamma$ 
& --- 
& $< 3.872 \left(95\%\right)$ 
& $< 3.934 \left(95\%\right)$  \\

$r_\mathrm{ur}$ 
& --- 
& $< 1.051 \left(95\%\right)$ 
& $< 1.123 \left(95\%\right)$  \\

$r$ 
& $< 1.207 \left(95\%\right)$ 
& $< 1.219 \left(95\%\right)$
& $< 1.252 \left(95\%\right)$ \\

$r_\mathrm{SM}$ 
& $< 1.241 \left(95\%\right)$ 
& $< 1.242 \left(95\%\right)$
& $< 1.252 \left(95\%\right)$ \\

\hline
\end{tabular}
\caption{Comparison of median and $1 \sigma$ values of cosmological parameters for $\Delta N_{\rm eff}$, decay model, and PT model scenarios.  For some parameters, 
we also present $95\%$ credible upper bounds.}
\label{tab:Posteriors}
\end{table}


\section{Conclusion}
\label{sec:Conclusion}

We have considered the general scenario in which energy is injected in the early 
Universe (either before or after BBN) in the form of both dark radiation and photons.  The 
impact of energy deposition on cosmological observables can be more complicated than the 
addition of extra effective neutrinos, since the addition of photon radiation can dilute the 
presence of dark radiation, leading to values of \DNeff which could be small, or even 
negative.

We first considered a scenario in which energy is present before BBN in the form of a 
scalar $Y$ that decays to dark radiation and electromagnetic radiation before recombination.
We find that in this scenario, even allowing an arbitrary mix of dark and electromagnetic radiation, 
constraints on the total amount of radiation which can be added are essentially no weaker than 
in the case in which only dark radiation is injected.  This is because there are tight constraints on 
the baryon-to-entropy ratio at BBN and at recombination, and an injection of electromagnetic energy after 
BBN would dilute the baryon density.  

However, in the case in which electromagnetic and dark radiation are added between BBN and 
recombination after a first-order phase transition, constraints on the total amount of radiation 
which can be injected are weakened marginally (by about $25\%$) compared to the case in which dark radiation 
alone is injected before BBN. This mild weakening of the constraints arises because the presence of 
dark radiation before BBN alters the Hubble parameter during BBN, affecting light element abundances.  That 
effect is absent if radiation is deposited after BBN as the result of a phase transition.

The ultimate reason why we have such tight constraints on the injection of radiation well before 
recombination is that we have considered a scenario in which the injected energy continues to redshift as 
radiation all the way to the epoch of recombination.  Since the matter densities (both CDM and baryonic) are 
tightly constrained at recombination, the injected radiation (whether dark or electromagnetic) will 
necessarily have some impact on cosmological observables.  The situation is considerably different if the 
dark radiation begins to redshift as matter well before recombination.  In this case, the dark radiation can 
just be considered a component of cold dark matter.  Since the cold dark matter density is not tightly constrained 
at BBN, the scenario in which some of the cold dark matter did not begin to redshift as matter until after BBN would 
be largely indistinguishable from $\Lambda$CDM.  However, if the injected energy only began to redshift as matter 
fairly close to the epoch of recombination, then the matter particles may not be sufficiently cold, and may contribute 
to the suppression of matter perturbations by free-streaming out of gravitational potentials.  As a result, there are 
bounds on Light (but Massive) Relics (LiMRs)~\cite{Xu:2021rwg} 
arising from observations of gravitational weak lensing of the CMB.  Nevertheless, 
even $\ev$-scale relics can contribute a non-trivial energy density, consistent with observations~\cite{Kumar:2025gkw}.

We have required the injection of energy to be effectively complete before the photons cool to a temperature of 
$10~\kev$, in order to ensure that the additional electromagnetic radiation can completely thermalize and not 
distort the CMB.  This assumption is made for simplicity only.  Photon deposition at later times may also be allowed, 
but would require a more detailed analysis to determine if the resulting CMB distortions are consistent with observation.  
A more detailed study of energy injection at later times would be an interesting topic of future work.

We have noted that our results are influenced by the mild discrepancy between the observed 
D/H ratio and that predicted by some nuclear reaction networks.  In particular, this discrepancy 
leads to a slightly lower prediction for the baryon-to-entropy ratio at BBN than the value obtained from 
the CMB.  This leads to an increased tension in any scenario in which entropy is injected between BBN and 
recombination.  It would be interesting to consider the extent to which constraints on the injection 
of entropy can be relaxed by using a different choice of nuclear reaction network.  More generally, we see 
the importance of resolving the discrepancies between the reaction networks.

{\bf Acknowledgements} We gratefully acknowledge Bhaskar Dutta and Jeremy Sakstein for useful discussions.  
JK and PS wish to acknowledge the Center for Theoretical Underground Physics and Related Areas (CETUP*), the Institute for Underground Science 
at Sanford Underground Research Facility (SURF), and the South Dakota Science and Technology Authority for hospitality and financial support, 
as well as for providing a stimulating environment.  JK would also like to acknowledge the University of Utah for its hospitality during the 
completion of this work.
JK is supported in part by DOE grant DE-SC0010504. PS is supported in part by National Science Foundation grant
PHY-2412834.

\appendix
\begin{widetext}
\section{Derivation of Comoving Energy Density Ratios}
\label{app:r_derivation}

We derive here analytic approximations for the quantities $r$, $r_\gamma$, and $r_\mathrm{ur}$ defined in Section~\ref{sec:GeneralScenario}.  We work in the early universe prior to $e^+e^-$ annihilation, when all SM species --- photons, electrons, positrons, and neutrinos --- are in thermal equilibrium at temperature $T_\gamma \approx 8 \mev$, and the dark sector particle $Y$ is also thermalized within the dark sector at temperature $T_\mathrm{dark} = \alpha \, T_\gamma$.

Before $e^+e^-$ annihilation, the SM radiation energy density is
\bea
\rho_\mathrm{SM} = g_{*\rho}^\mathrm{pre} \frac{\pi^2}{30} T_\gamma^4 ,
\eea
where $g_{*\rho}^\mathrm{pre} = 2 + (7/8)(4 + 6) = 43/4$ counts photons ($g=2$), electrons and positrons ($g=4$), and three neutrino flavors ($g=6$).

The energy density of the relativistic spin-0 dark sector particle $Y$ is
\bea
\rho_Y = \frac{\pi^2}{30} T_\mathrm{dark}^4 = \frac{\pi^2}{30} \alpha^4 T_\gamma^4 .
\eea
To express the $Y$ energy density in terms of an effective neutrino contribution, we compare to the energy density of a single SM neutrino species at the same temperature $T_\gamma$ (before neutrino decoupling):
\bea
\rho_{1\nu} = \frac{7}{8} \cdot 2 \cdot \frac{\pi^2}{30} T_\gamma^4 
= \frac{7}{4}  \frac{\pi^2}{30} T_\gamma^4.
\eea
Then
\bea
\DNeff^\mathrm{BBN} \equiv \frac{\rho_Y}{\rho_{1\nu}} 
= \frac{\alpha^4}{7/4} = \frac{4}{7} \alpha^4 ,
\eea
as quoted in Eq.~\ref{dneff_bbn} of the main text.

The ratio $r$ compares the total comoving radiation energy density at a reference time $t_f$ after both $e^+e^-$ annihilation and the decay of $Y$ have completed to that at $t_i$ before $e^+e^-$ annihilation, when $Y$ is still relativistic:
\bea
r \equiv \frac{a_f^4 \rho_r^f}{a_i^4 \rho_r^i} .
\eea
The initial comoving radiation energy density, expressed in terms of the pre-annihilation photon comoving density $\bar\rho_\gamma^i$, is
\bea
\bar\rho_r^i \equiv a_i^4 \rho_r^i 
= \bar\rho_\gamma^i \frac{g_{*\rho}^\mathrm{pre} + \alpha^4}{2}
\nonumber \\ 
= \bar\rho_\gamma^i 
\left[\frac{43}{8} + \frac{7}{8}\DNeff^\mathrm{BBN}\right] ,
\eea
where the first equality follows from all species sharing temperature $T_\gamma$ before annihilation, so that $\rho_r^i/\rho_\gamma^i = (g_{*\rho}^\mathrm{pre} + \alpha^4)/2$, with the $\alpha^4/2 =(7/8)\DNeff^\mathrm{BBN}$ contribution from $Y$.  We then have
\bea
r &=& \frac{\bar\rho_r^f}{\bar\rho_r^i} = \frac{\bar\rho_\gamma^f + \bar\rho_\mathrm{ur}^f  + \bar\rho_\mathrm{ncdm}^f}{\bar\rho_\gamma^i + \bar\rho_\mathrm{ur}^i + \bar\rho_\mathrm{ncdm}^i + \bar\rho_Y^i} = \frac{\bar\rho_\gamma^f + \bar\rho_\mathrm{ur}^f + \bar\rho_\mathrm{ncdm}^f}{\bar\rho_\gamma^i 
\left[\frac{43}{8} + \frac{7}{8}\DNeff^\mathrm{BBN}\right] } .
\eea
In practice, we compute $r$, $r_\gamma$, and $r_\mathrm{ur}$ numerically by solving 
the Boltzmann equations (Eq.~\ref{eq:Boltzmann}) directly, using \texttt{LINX}.

\subsection*{Phase transition case}
For the phase transition scenario described in Section~\ref{sec:GeneralScenario}, the energy injection occurs after BBN, so there is no contribution to $\DNeff^\mathrm{BBN}$ from the dark sector at the time of nucleosynthesis. The phase transition injects energy in form of both photons and dark radiation, increasing the photon and ultra-relativistic comoving energy densities by factors $r_\gamma$ and $r_\mathrm{ur}$, respectively.

The initial comoving radiation density (before the phase transition and before $e^+e^-$ annihilation) is
\bea
\bar\rho_r^i = \bar\rho_\gamma^i \frac{g_{*\rho}^\mathrm{pre}}{2} =  \frac{43}{8} \bar\rho_\gamma^i
\eea
where we have included only SM photons and neutrinos, with no dark sector contribution at BBN.

After the phase transition and $e^+e^-$ annihilation, the photon and ultra-relativistic particle comoving densities are
\bea
\bar\rho_\gamma^f &=& r_\gamma \bar\rho_\gamma^i , \nonumber\\
\bar\rho_\mathrm{ur}^f &=& r_\mathrm{ur} \bar\rho_\mathrm{ur}^i =  r_\mathrm{ur}  \frac{14}{8} \bar\rho_\gamma^i.
\eea
And the ratio of total comoving radiation densities is
\bea
r &\equiv& \frac{\bar\rho_r^f}{\bar\rho_r^i} 
= \frac{r_\gamma \bar\rho_\gamma^i + \dfrac{14}{8} r_\mathrm{ur} \bar\rho_\gamma^i + \dfrac{7}{8}  \bar\rho_\gamma^i  }{\dfrac{43}{8} \bar\rho_\gamma^i }
= \frac{r_\gamma + \dfrac{14}{8} r_\mathrm{ur}  + \dfrac{7}{8}   }{\dfrac{43}{8} } .
\eea
In the case where no energy is injected from the phase transition, we have  $r_\gamma= \left(11/4\right)^{4/3},r_\mathrm{ur} =1$, and we recover the \lcdm value 
$r_{\Lambda\mathrm{CDM}} = \left[\left(11/4\right)^{4/3} +  21/8 \right]/ \left[43/8\right] \approx 1.205$.
\end{widetext}

\section{Complete Posteriors}
\label{sec:CompletePosteriors}

\begin{figure*}
    \centering
    \includegraphics[width=0.99\linewidth]{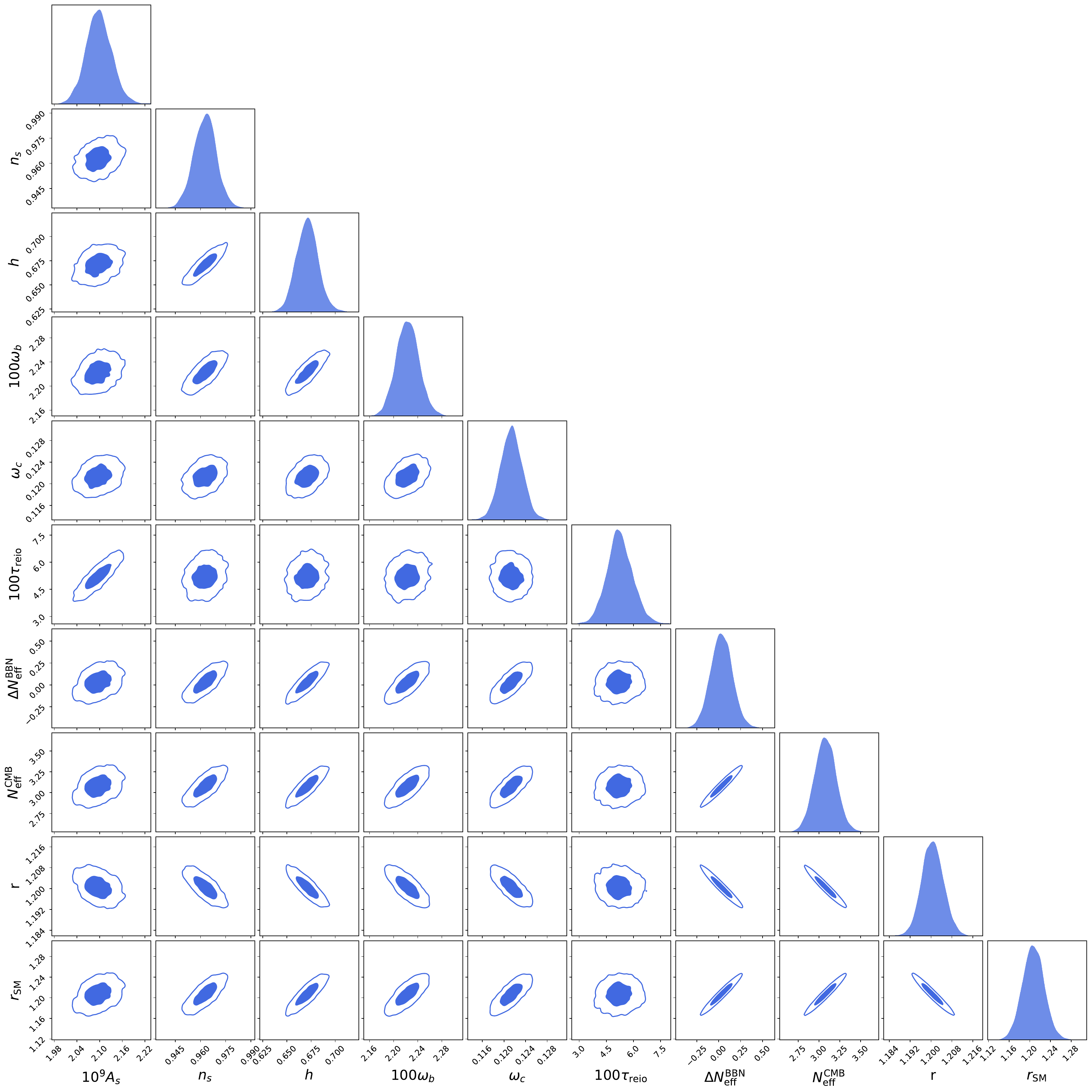}
    \caption{Posterior distributions for the $\Delta N_\mathrm{eff}$ model parameters and derived quantities, obtained from a joint analysis of \textit{Planck} 2018 CMB temperature and polarization anisotropies combined with primordial $^4$He and D/H abundance measurements. BBN predictions are computed using \texttt{LINX} with the PRIMAT nuclear reaction network. Contours show $68\%$ and $95\%$ credible intervals.}
    \label{fig:cmb_bbn_neff_results}
\end{figure*}

\begin{figure*}
    \centering
    \includegraphics[width=0.99\linewidth]{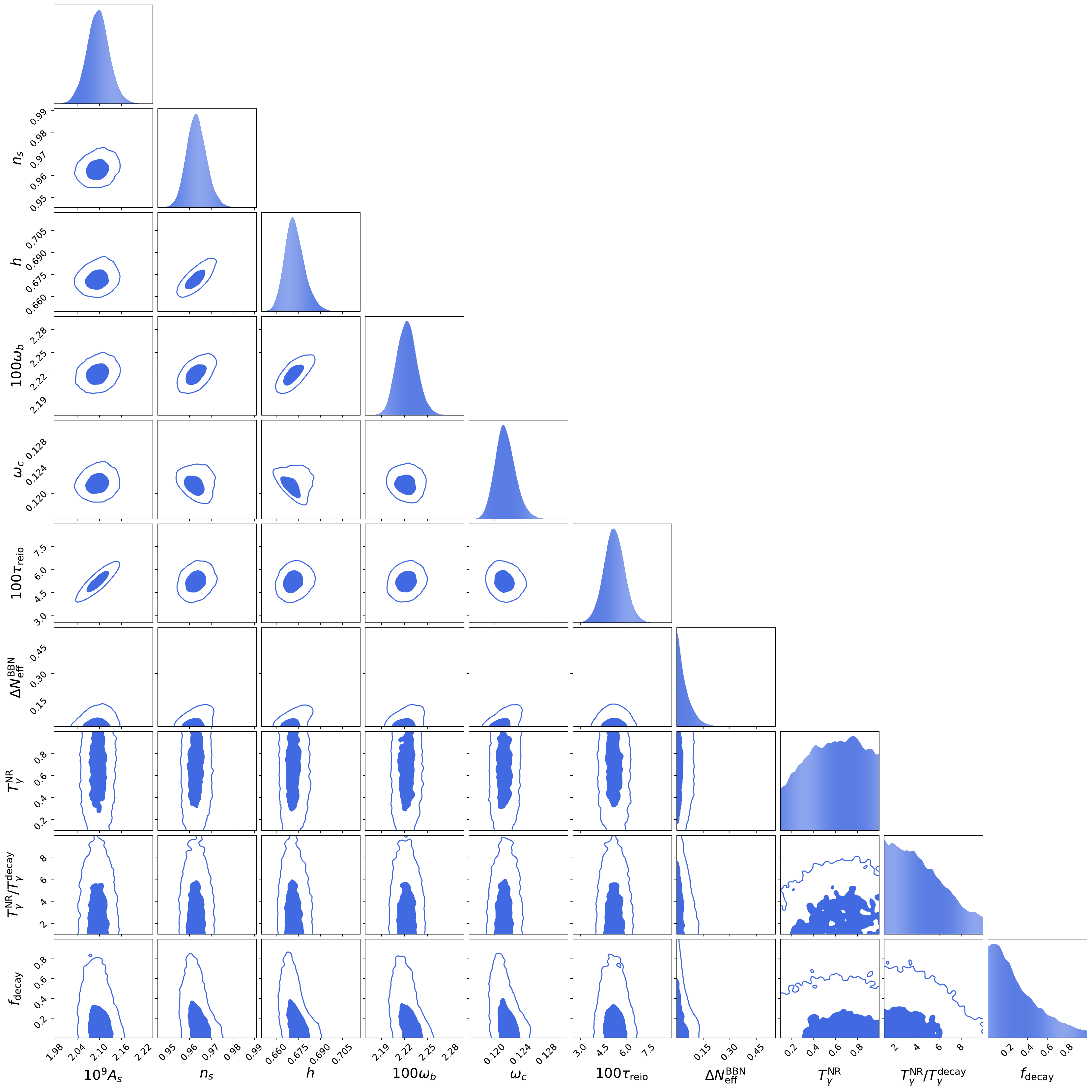}
    \caption{Posterior distributions for the decay model parameters and derived quantities, obtained from a joint analysis of \textit{Planck} 2018 CMB temperature and polarization anisotropies combined with primordial $^4$He and D/H abundance measurements. BBN predictions are computed using \texttt{LINX} with the PRIMAT nuclear reaction network. Contours show $68\%$ and $95\%$ credible intervals.}
    \label{fig:cmb_bbn_decay_results_appendix}
\end{figure*}

\begin{figure*}
    \centering
    \includegraphics[width=0.99\linewidth]{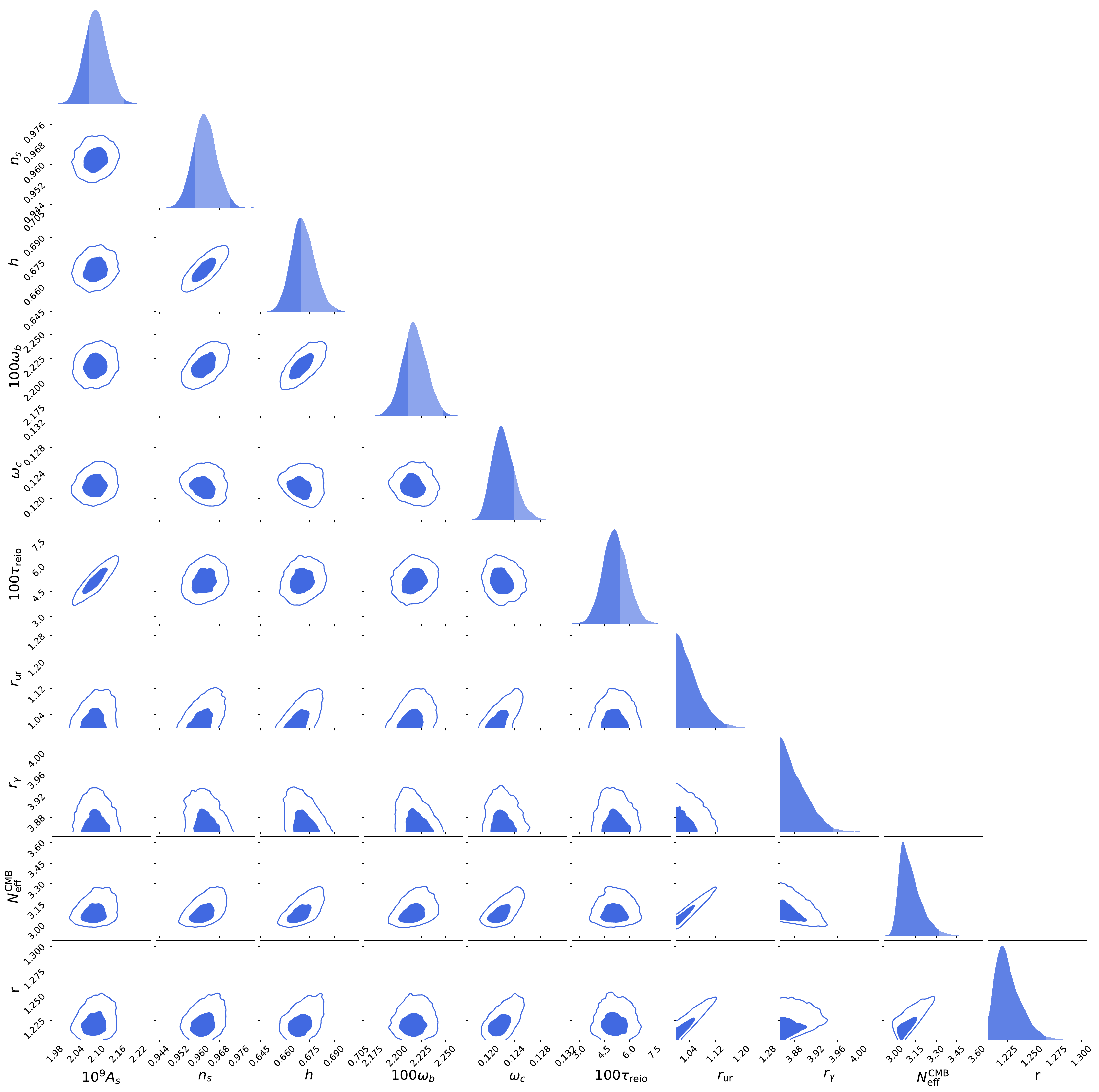}
    \caption{Posterior distributions for the phase transition (PT) model parameters and derived quantities, obtained from a joint analysis of \textit{Planck} 2018 CMB temperature and polarization anisotropies combined with primordial $^4$He and D/H abundance measurements. BBN predictions are computed using \texttt{LINX} with the PRIMAT nuclear reaction network. Contours show $68\%$ and $95\%$ credible intervals.}
    \label{fig:cmb_bbn_PT_results}
\end{figure*}

In this section, we present the complete posteriors for the $\Delta N_\mathrm{eff}$ model (Figure~\ref{fig:cmb_bbn_neff_results}), 
the decay model (Figure~\ref{fig:cmb_bbn_decay_results_appendix}) and the PT model (Figure~\ref{fig:cmb_bbn_PT_results}).  The 
additional parameters are described in Section~\ref{sec:Analysis}.


\end{document}